\documentstyle[prl,aps,epsf,float,twocolumn]{revtex}
\begin{document}
\draft

\twocolumn[\hsize\textwidth\columnwidth\hsize\csname@twocolumnfalse\endcsname
\title{Soundwave Anomalies in SrCu$_{2}$(BO$_{3}$)$_{2}$}
\author{S.Zherlitsyn$^{(1)}$, S.Schmidt$^{(1)}$, B.Wolf$^{(1)}$,
H.Schwenk$^{(1)}$, B.L\"{u}thi$^{(1)}$,\\ H.Kageyama$^{(2)}$,
K.Onizuka$^{(2)}$, Y.Ueda$^{(2)}$, K.Ueda$^{(2)}$}
\address{$^{(1)}$Physikalisches Institut, Universit\"{a}t Frankfurt,
                 D-60054 Frankfurt, Germany\\
         $^{(2)}$Institute for Solid State Physics, University of Tokyo,
Roppongi 7 - 22 - 1, Tokyo 106 - 8666, Japan  }
\date{\today}
\maketitle

\begin{abstract}
The temperature and high magnetic field dependence of the longitudinal
soundwave mode $c_{11}$ in the two dimensional dimer system
SrCu$_{2}$(BO$_{3}$)$_{2}$ is presented. $c_{11}(T)$ shows anomalies due to
strong interdimer spin-strain coupling. We can quantitatively interpret the
temperature dependence of $c_{11}(T)$ together with the magnetic susceptibility
$\chi_{m}(T)$ with a molecular field approximation of coupled dimer triplets.
The sound velocity up to 50T shows very sharp softening between the
magnetization plateaus at low temperature. We argue that these pronounced
effects arise from a resonant interaction between the phonons and the magnetic
excitations which show softening between the plateaus.
\end{abstract}
\pacs{PACS numbers: 62.65.+k, 75.10.Jm, 75.40.Cx } ]
\newpage

The physics of low dimensional spin systems is full of surprises. In zero
magnetic field one finds gapped ground states for antiferromagnetic Heisenberg
chains with integer spins \cite{Haldane} or for dimerised spin 1/2 systems
(e.g. spin-Peierls systems \cite{spin-Peierls}) or for spin 1/2 two leg ladders
\cite{Dagotto} to name just a few examples. Also in the presence of a magnetic
field instead of a smooth magnetization curve one has found for some classes of
materials \cite{Totsuka} magnetization plateaus for rational fractions of the
saturation magnetization. Examples given are the quasi one dimensional
Ni-compounds \cite{Sato} , NH$_{4}$CuCl$_{3}$ \cite{Shiramura} and the two
dimensional SrCu$_{2}$(BO$_{3}$)$_{2}$ \cite{Kagayama}.

\begin{figure}
\centerline{\epsfxsize=3.3in\epsffile{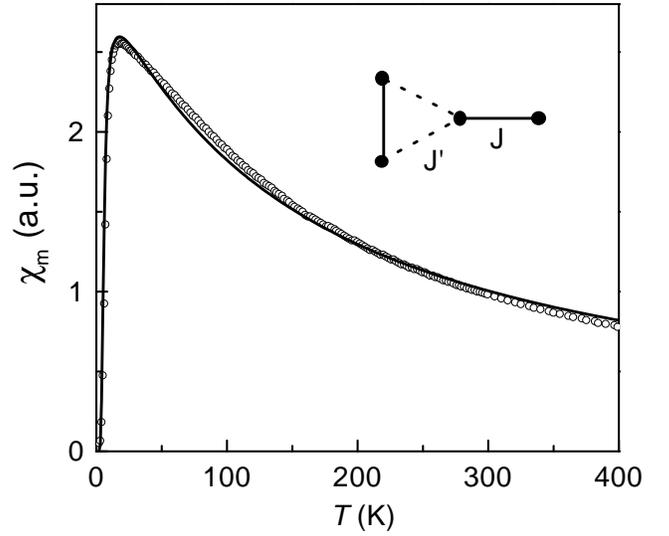}} \caption{Temperature
dependence of magnetic susceptibility. Open circles: experiment from reference
[7], full line: fit with expression equation (1). Inset: two neighboring dimers
with exchange interactions $J,J'$.}\label{fig1}
\end{figure}

This latter compound has a two dimensional structure which consists of
alternately stacked  CuBO$_{3}$- and Sr- layers. Within the CuBO$_{3}$ plane
the Cu$^{2+}$- Cu$^{2+}$ dimers are orthogonally connected giving rise to an
exact orthogonal dimer ground state \cite{Miyahara99}. This system is
topologically equivalent to the Shastry-Sutherland model \cite{Shastry}. In the
inset of figure \ref{fig1} two orthogonal dimers with the important exchange
interactions $J, J'$ are shown. Numerical evaluation of the corresponding
Hamiltonian gives $J$ = 100K, $J'$ = 68K \cite{Miyahara99,Miyahara00,Weihong}.
These authors also noted that the value $J'/J$ = 0.68 is very close to the
value of a quantum phase transition $(\approx 0.7)$.

In this letter we report the following new features for this compound: (1) The
longitudinal elastic constant $c_{11}$ exhibits a strong temperature dependence
which allows to study the strain dependence and the elastic coupling between
the Cu-dimers. An important result of this analysis leads to an estimate for a
possible structural phase transition. (2) The sound velocity in high magnetic
fields up to 50T exhibits characteristic features related to the magnetization
plateaus, very sharp minima before successive plateaus. A resonant interaction
between the soundwave and the gap modes which soften between the plateaus might
be responsible for these anomalies. This evidence is a direct proof for the
softening of the magnetic excitation between the plateaus. A theoretical
treatment for these effects especially a calculation of the magnetic excitation
is lacking so far.

\begin{figure}
\centerline{\epsfxsize=3.3in\epsffile{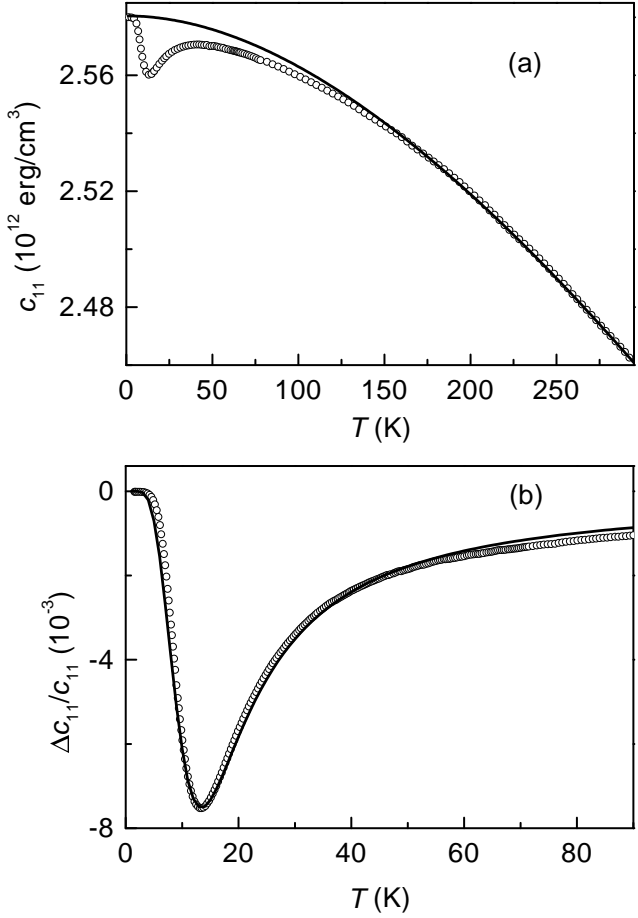}} \caption{(a) Temperature
dependence of the $c_{11}$mode. Full line is background $c_{11}^{o}(T)$ as
described in the text. The ultrasonic frequency was 98MHz.  (b) $(c_{11}(T) -
c_{11}^{o}(T))/c_{11}^{o}(T=0)$ as a function of temperature. Open circles:
experiment, full line: fit from eq.(2).} \label{fig2}
\end{figure}

In figure \ref{fig1} and \ref{fig2} we show the magnetic susceptibility and the
elastic constant $c_{11}$ as a function of temperature. Both thermodynamic
quantities exhibit features typical for a low dimensional spin gap system.
The magnetic susceptibility $\chi_{m}$, published before
\cite{Kagayama,Miyahara99} has a maximum at $T_{max}$ = 18K and a strong
decrease for lower temperatures indicative for a spin gap system. At high
temperatures a Curie-Weiss behaviour with $\Theta = (J+4J')/4 =$ 93K is
observed. For the following it is interesting to note that a RPA-molecular
field expression \cite{White}
\begin{equation}\label{RPAmagnetik}
\chi_{m} = \chi_{o}/(1 - j \chi_{o})
\end{equation}
gives a rather good fit for $T <$ 300K. $\chi_{0}$ is the magnetic
susceptibility of an isolated dimer, a singlet - triplet system:  $\chi_{0} =
2e^{-\Delta / kT}/kTZ$ with $Z$ the partition function. The fit gives a singlet
- triplet gap for thermodynamic quantities $\Delta$ = 29K rather close to the
accepted value of 30K\cite{Miyahara99} and $j$ = -273K. The inter dimer
coupling is $j = 4J'$ as expected \cite{Miyahara99}. Instead of starting with
the Hamiltonian with the parameters $J, J'$ \cite{Miyahara99} we used directly
the dimer susceptibility $\chi_{0}$ with the renormalized value for $\Delta$ =
29K $< J$ = 100K.
In figure \ref{fig2} we show the temperature dependence of the longitudinal
constant $c_{11}$. For this mode the polarization and the propagation is along
[100]. With the sound velocity $v$ and the mass density $\rho=$4.1g/cm$^{3}$
one gets $c_{11}=\rho v^{2}$. In figure \ref{fig2}(a) we plot $c_{11}(T)$
together with the temperature dependent background $c_{11}^{0}(T)$ taken from
an empirical evaluation \cite{Varshni}. Similar to the magnetic susceptibility
$\chi_{m}$ we can interpret the temperature dependence of the elastic constant
$c_{11}$ with an analogous RPA expression \cite{Luethi}. In analogy to the
magnetic susceptibility which is the response function of the magnetization to
an applied field we can define the strain susceptibility which is the response
function to an applied strain.The elastic constant reads:
\begin{equation}\label{RPAstrain}
\begin{array}{c}
c_{11}(T) = c_{11}^{o}(T) - G^{2}\chi_{str}\mbox{  with} \nonumber \\
\chi_{str}=\chi_{s}/(1-K\chi_{s})
\end{array}
\end{equation}
$\chi_{s}$ is the strain susceptibility of a single dimer
\begin{equation}\label{chistrain}
\chi_{s} = \frac{\partial^{2}F}{\partial\epsilon^{2}} = \frac{3e^{-\Delta/ kT
}}{kTZ^{2}}
\end{equation}
and $K$ is the strength of the ($k=0$) dimer - dimer interaction. The effective
interaction $K$ can be mediated e.g. by phonons.

In figure \ref{fig2}(b) we compare $\Delta c_{11}/c_{11}$ $ = (c_{11}(T) -
c_{11}^{o}(T))/c_{11}^{o}(T=0)$ with a fit to $\chi_{str}$ using the same
$\Delta$ = 29K. Here the minimum of the elastic constant is at $T_{min}$ = 13K
and the coupling constant $K$ = 23.8K. The fit is also very good. This analysis
shows that with a 3 times larger coupling constant $K$ a structural transition
would take place. The single dimer coupling constant $G =
\partial\Delta/\partial\epsilon$ which measures the strain dependence of the singlet -
triplet gap amounts to $G \approx$ 900K (with $c^{o}(T=0)$ = $25.8\cdot10^{11}$
erg/cm$^{3}$ at low temperature and the number of dimers $n =
0.74\cdot10^{22}$cm$^{-3}$). So this simple RPA expressions account correctly
for the $T_{max}$ and $T_{min}$ in $\chi_{m}$ and $\chi_{str}$ respectively. It
also gives a good overall fit for $\chi_{m}$ and $c_{11}$ using the same
singlet - triplet gap. It further shows that the strain interactions are very
important for this dimer compound.
\begin{figure}
\centerline{\epsfxsize=3.3in\epsffile{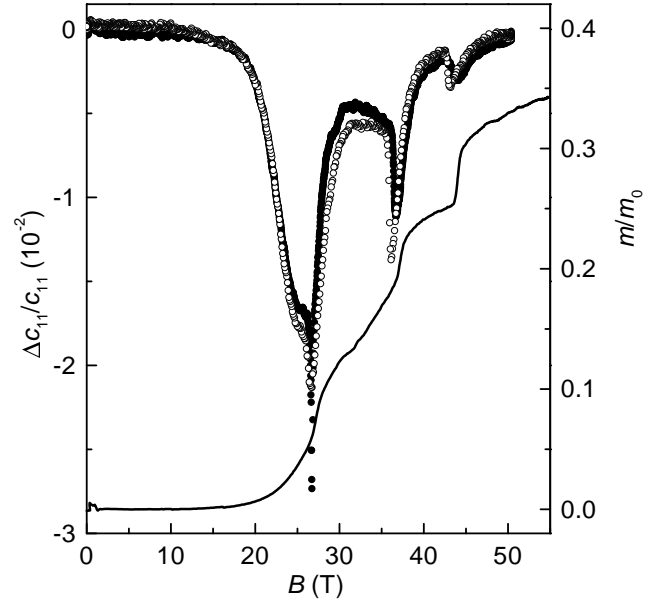}} \caption{Relative change
of the elastic constant  $c_{11}$ for the magnetic field  sweep up (solid
circles) and  sweep down (open circles) measured at $T = 1.5$K. Field
dependence of relative magnetization $m/m_{0}$ at the same temperature is also
shown (solid line). Here $m_{0}$ is a saturation value of the magnetization.
The magnetic field is applied along the $a$-axis. The elastic constant minima
occur between the magnetization plateaus. } \label{fig3}
\end{figure}
Now we proceed to the magnetic field dependence. In figure \ref{fig3} we show
the relative change of the same elastic constant $c_{11}$ as a function of
magnetic field up to 50T at 1.5K. The ultrasonic technique recently developed
for pulsed fields is described elsewhere \cite{Wolf}. In the present case the
rise time of the pulse is 8ms and the total pulse duration is 24ms. In figure
\ref{fig3} both curves for field up and down are shown. No substantial heating
effects are discernible, we estimate it to $\Delta T <$ 0.5K. For 1.5K we
observe a broad minimum at 25T followed by very sharp minima at 27T, 36T and
42T. In this figure the magnetization for the same temperature is also
included. It clearly exhibits magnetization plateaus for $m/m_{0} =$ 1/8, 1/4
and 1/3. It is seen that the sound velocity minima are in the region where the
magnetization changes from one plateau to the next one.

The broad minimum at 25T arises possibly  from localised triplet excitations.
Existence of a strong spin-phonon coupling was pointed out recently by Ueda and
Miyahara \cite{Ueda}. If the spin-phonon coupling is strong enough there is the
possibility of the formation of a self-trapped triplet, a bound complex of
triplet excitation and local distortion (triplet-phonon-bound state). The
energy of the self trapped triplet excitations is lower than the usual triplet
branch, which may explain this broad minimum.
\begin{figure}
\centerline{\epsfxsize=3.3in\epsffile{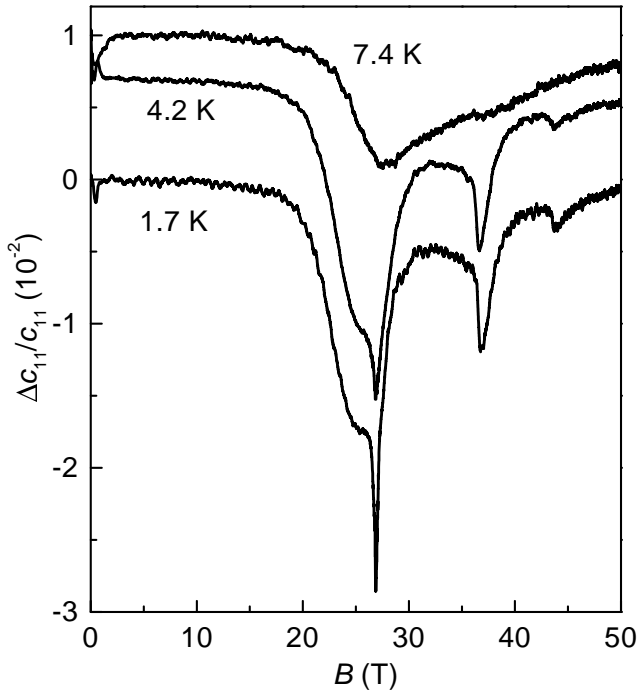}} \caption{Relative change
of the elastic constant  $c_{11}$ versus magnetic field at three different
temperatures. Only field sweeps up are shown. The curves were shifted
arbitrarily for the sake of clarity. } \label{fig4}
\end{figure}
The first strong softening at 27T is due to the interaction between the
soundwave and the soft triplet branch. Inelastic neutron scattering determined
$\Delta$ = 35K \cite{Kageyama to be publ}. This leads to a softening at 27T
with $g = 2$ in agreement with our observation. The same effect we observed
recently in (VO)$_{2}$P$_{2}$O$_{7}$ \cite{Wolf2000}. The evolution of this
sound velocity minimum with temperature is the same in both substances:
Comparing $\Delta c_{11}/c_{11}$ for 1.7K and 7.4K we observe a shift to higher
fields and a broadening with increasing temperature. This is seen in figure
\ref{fig4} where the field dependence of the relative elastic constant is
plotted for different temperatures. The sharp minima between the plateaus
disappear rapidly with increasing temperature and at 7.4K only the first
minimum survives.

We can estimate again the coupling constant $\partial\Delta/\partial\epsilon$
from the softening of the first triplet at 27T where $\Delta c_{11}/c_{11}
\approx 2.8\%$. Since the first magnetic excitation shows no dispersion
\cite{Kageyama to be publ} we can again use the free energy of weakly coupled
dimers. We get $\partial\Delta/\partial\epsilon \approx$ 700K in rough
agreement with the estimate from the $c_{11}$ temperature dependence given
above. This indicates that both effects are due to an exchange striction
coupling.

Now we discuss the additional steep minima effects at 36T and 42T, the steep
minima before each plateau ($m/m_{0}=1/4, 1/3$). We argue that these are
likewise resonant effects between the sound wave and the corresponding soft
mode before the plateau is reached. Within the plateau the excitation modes
have a gap, thus keeping the magnetization constant. While we are not aware of
any theory describing these excitation modes and its interaction with the
elastic modes we can give the field dependence  of our sound velocities between
the plateaus: $ \Delta c_{11}/c_{11} \propto (B-B_{c})^{\gamma} $ with $ \gamma
\approx 0.5 $ for the resonances before the plateaus for $B>B_{c}$ at $m/m_{0}=
1/4, 1/3$.

In summary we have given temperature and field dependencies of a longitudinal
elastic constant in the two dimensional compound SrCu$_{2}$(BO$_{3}$)$_{2}$. We
find considerable phonon induced dimer - dimer coupling and a strain dependence
of the singlet - triplet gap of 700-900K in rough agreement with exchange
striction coupling constants ($\partial J/ \partial \epsilon \approx -9J$)
\cite{Luethi}. The sharp softening of $c_{11}$ between the plateaus indicates a
resonant interaction with the magnetic excitation spectrum which softens
between the plateaus. Further investigations of other elastic modes can help in
elucidating the strain - magnetic excitation interaction in this material. Is
is clear that phonon effects are very important in this compound. A more
detailed theoretical treatment of this interaction and of the magnetic
excitations is needed.

This research was supported in part by a grant from BMBF 13N6581A/3 and by a
Grant-in-Aid for Encouragement Young Scientists from the Ministry of Education,
Science, Sports and Culture, Japan. A helpful discussion with G.S.Uhrig is
gratefully acknowledged.

\end{document}